\documentclass[fleqn,twoside]{article}
\usepackage{espcrc2}
\usepackage{graphicx,epsfig,rotating}

\newcommand{\lsim}   {\mathrel{\mathop{\kern 0pt \rlap
  {\raise.2ex\hbox{$<$}}}
  \lower.9ex\hbox{\kern-.190em $\sim$}}}
\newcommand{\gsim}   {\mathrel{\mathop{\kern 0pt \rlap
  {\raise.2ex\hbox{$>$}}}
\lower.9ex\hbox{\kern-.190em $\sim$}}}

\def\be{\begin{equation}}
\def\ee{\end{equation}}
\def\ba{\begin{eqnarray}}
\def\ea{\end{eqnarray}}

\title{Ultra High Energy Neutrino Astronomy }

\author{V. Berezinsky\address[LNGS]{INFN - Laboratori Nazionali del Gran Sasso,
        I--67010 Assergi (AQ), Italy}}

\begin{document}

\begin{abstract}
The short review of theoretical aspects of ultra high energy (UHE) neutrinos 
and  superGZK neutrinos. The sources and diffuse fluxes of UHE
neutrinos are discussed. Much attention is given to comparison of the 
cascade and cosmic ray upper bounds for diffuse neutrino fluxes. 
Cosmogenic neutrinos and neutrinos from the mirror mater are 
considered as superGZK neutrinos. 

\end{abstract}

\maketitle
\section{Introduction}

With Baikal and AMANDA detectors UHE neutrino astronomy entered its
first stage, which will reach the decisive level with 1~km$^3$ detector. 
Simultaneously, the first steps are made in building the detectors 
searching for superGZK neutrinos, i.e. those with energies above 
the GZK cutoff, $E_{\rm GZK} \sim 5\times 10^{19}$~eV. 
Radio-telescopes, such as RICE, GLUE and FORTE are already 
searching for radio signal from neutrino-induced showers in ice and in 
the moon. Atmospheric neutrino-induced showers can be observed from
the space
as it is proposed in the EUSO project. Neutrino-induced horizontal atmospheric 
showers can be observed by the Auger array, which is now in
the data-collecting phase. \\*[2mm]
{\em Production of UHE cosmic neutrinos}\\ 
occurs in $pp$ and $p\gamma$
collisions of UHE protons with the target protons/nuclei and with target 
low-energy photons. They can be also produced by annihilation 
of DM particles and by decays of superheavy particles. In all
these cases neutrinos are mostly produced  in the chain of  
pion decays, and hence neutrino astronomy is a search for locations 
of HE pion production.\\*[2mm] 
{\em UHE neutrino sources}\\ 
can be subdivided into accelerator sources
and top-down sources, where neutrinos are produced in decays 
and annihilation of heavy particles. The examples of such sources are 
given by annihilation of neutralinos in the Sun and Earth, by
topological defects, which produce superheavy unstable particles, 
and by decays of quasi-stable superheavy DM particles. 

Usually neutrino radiation is accompanied by other types of HE radiations,
most notably by HE gamma-radiation and cosmic rays (CR). There are, however,
so called ``hidden sources'' where all accompanying radiations are 
strongly or almost absorbed. The examples of such objects are the Sun
and Earth, in center of which neutralinos annihilate. Another ideal example  
is given by mirror matter, where all mirror particles interact with 
visible matter gravitationally, and only mirror neutrinos can
oscillate into visible ones. The almost ``hidden'' source is given by 
the Stecker model \cite{stecker} of AGN, where UHE photons and
protons are mostly absorbed or confined, and only HE neutrinos emerge
from there. There are some other less exotic examples of astrophysical 
hidden sources \cite{book}.\\*[2mm] 
\noindent
{\em Neutrino detection}\\ 
includes four remarkable reactions:\\*[1mm]
\noindent
Muon production $\nu_{\mu}+N \to \mu+ {\rm all}$ gives an excellent tool
to search for the discrete sources, since directions of UHE muon and 
neutrino coincide. \\*[1mm]
\noindent
Resonant production of W-boson, $\bar{\nu}_e + e \to W^-\to {\rm hadrons}$ 
results in production of monoenergetic showers with energy 
$E_0=m_W^2/2m_e=6.3\times 10^6$~GeV. This reaction has a large cross-section.
\\*[1mm]
\noindent
Tau production in a detector, $\nu_{\tau}+N \to \tau + {\rm hadrons}$,
is characterised by time sequence of three signals: a shower from prompt
hadrons, the Cherenkov light from $\tau$ and hadron shower from $\tau$-decay.
SuperGZK $\nu_{\tau}$ are absorbed less in the Earth due to
regeneration: absorbed $\nu_{\tau}$ is converted into $\tau$, which decays
producing $\nu_{\tau}$ again. \\*[1mm]
\noindent
Z-bursts provide a signal from the space, caused by the 
resonant $Z^0$ production on 
DM neutrinos, $\nu+ \bar{\nu}_{\rm DM} \to Z^0 \to {\rm hadrons}$.
The energy of the detected neutrino must be tremendous:
\begin{equation}
E_0=\frac{m_Z^2}{2m_{\nu}}=1.7\times 10^{13}\left (\frac{0.23~{\rm eV}}
{m_{\nu}}\right )~{\rm GeV}.
\label{Zburst}
\end{equation}
\noindent
{\em Neutrino oscillations}\\ 
play the essential role. 
The neutrino flavors $\bar{\nu}_e$ and $\nu_{\tau}$ are inefficiently
produced in the accelerator sources. The flavor oscillation with 
probability
$$
P_{\nu_{\alpha}\to \nu_{\beta}}=\frac{1}{2}\sin^22\theta
\sin^2\frac{r}{L(E)},
$$
and oscillation length $L(E)$, given by
$$
L(E)=\frac{4E}{\Delta m^2}=25\left ( \frac{E}{10^{11}~{\rm GeV}}\right )
\left (\frac{10^{-4}{\rm eV}^2}{\Delta m^2}\right )~{\rm pc}
$$
are very efficient due to large mixing angles $\theta$ and
astrophysically very short oscillation lengths. 
If initial flavor ratio is given by pion decays, 
$\nu_e:\nu_{\mu}:\nu_{\tau}=1:2:0$, the observed flavor ratio is 
$\nu_e:\nu_{\mu}:\nu_{\tau}=1:1:1$ (equipartion).\\*[2mm]
{\em HE neutrinos from early universe}.\\
One might think (and many did think) that large neutrino fluxes can 
be produced at cosmological epochs with large red shift. This
possibility is disfavored \cite{nucl} by absorption of HE neutrinos and by
nucleosynthesis bound on their fluxes. Neutrinos are absorbed in
$\nu\bar{\nu}$ collisions with big-bang neutrinos and horizon of 
observation for neutrinos with energy $E_{\nu0}$ (at present) is given
by redshift
$$
z_{\rm abs}=7.9\times 10^4 (E_{\nu 0}/{\rm 1~TeV})^{-1/3}
$$
Neutrino fluxes produced at large z are 
strongly restricted by production of $D$ and $^3He$ at the epochs 
after Bing Bang Nucleosynthesis. Neutrinos cause e-m cascades and 
MeV photons from these cascades produce $D$ and $^3He$ in
collisions with $^4He$ nuclei:    
$$
\nu+\bar{\nu}_{\rm DM}\to {\rm cascade} \to {\rm MeV~photons} \to\\
$$
$$
\gamma_{\rm MeV}+^4\!{\rm He} \to ^3\!\!{\rm He/D}.
$$
\section{Accelerator sources}
Protons are assumed to be accelerated mostly by the shocks and produce 
neutrinos in pp-collisions in gas and in $p\gamma$ collisions in
low-energy photon background.\\ 
{\em HE neutrinos from AGN cores.}\\*[1mm]
This case is known as the Stecker model \cite{stecker}. The massive
black hole (BH) is surrounded by the thick accretion disc. The radial flow
of accreting gas to BH is terminated by the shock at radial distance 
$r_{\rm sh}\sim 10 r_g$. The protons are accelerated diffusively at the shock, 
but they cannot diffuse upstream. As a result they are dragged by gas
flow downstream and are absorbed by BH.  Neutrinos are produced in 
$p\gamma$-collisions with UV photons of the disc radiation. 
HE gamma produced in the same  $p\gamma$-collisions are absorbed 
due to $\gamma+\gamma_{\rm UV} \to e^++e^-$. Neutrino flux from an 
individual AGN is not detectable, while the diffuse flux is detectable.    
In fact, the original prediction of the Stecker model is higher than
upper limit of AMANDA, but diffuse flux in the Stecker model can be easily
suppressed by the choice of parameters, in particular by increasing of 
$r_{\rm sh}$. \\*[2mm]
{\em HE neutrinos from AGN jets} \cite{jet}.\\*[1mm]
\noindent
The protons are
accelerated in the multiple shocks in the AGN jets, especially in 
their inner parts. Neutrinos are produced in the collisions with
photons from the accretion disc and from photons produced in the jet 
by accelerated electrons and photons. Neutrino flux from individual 
AGN is very small, but the diffuse flux is detectable by future 
1~km$^3$ detector.\\*[2mm]   
{\em HE neutrinos from galaxy clusters} \cite{cluster}.\\*[1mm]
The clusters of galaxies are able to keep UHE cosmic rays for a time 
exceeding the age of the universe. This is the key phenomenon which 
makes galaxy clusters the powerful sources of UHE neutrinos. The
particles are accelerated in clusters by various mechanisms: 
in the normal galaxies by SN shocks, in AGN and cD-galaxies, in 
the process of galactic merging. HE protons, confined in a cluster by 
intracluster magnetic fields, produce the large flux of $pp$-neutrinos due to
long confinement time. The diffuse HE neutrino flux is determined
entirely by basic parameters characterising the clusters. In particular,
for the lower limit of the diffuse flux provided  by normal galaxies  
with CR luminosity $L_p$ and generation index $\gamma_g$, both taken
as ones in our galaxy, the diffuse flux is given as 
$$
J_{\nu}(E) \propto L_p E^{-\gamma_g}\frac{N_g}{R_{\rm cl}^3}\xi 
\Omega_b\rho_{\rm cr},
$$
where $R_{\rm cl} \sim 2$~Mpc is the virial radius of a cluster, 
$N_g \sim 100$ is richness of a cluster, $\rho_{\rm cr}$ is critical
cosmological density, and $\xi\Omega_b$ is cosmological baryonic density
provided by clusters. \\*[2mm]
{\em Gamma Ray Bursts}.\\*[1mm]
\noindent
GRBs are most exiting sources of UHE neutrinos. There are two
mechanisms of HE neutrino generation. In the first one \cite{vietri} 
particles are accelerated by external shock and neutrinos are produced 
in $p\gamma$ collisions with GRB photons behind the shock. In the
second mechanism \cite{waxman} protons are accelerated by internal shocks. 
The shocks collide and produce the turbulence with equipartion
magnetic field. In such medium the Fermi II acceleration mechanism operates,
the spectrum is assumed to be $\propto 1/E^2$. Neutrinos are produced in 
$p\gamma$ collisions with GRB photons. All estimates are very
transparent and follow from assumption that the energy outputs 
in GRB photons, accelerated protons and produced neutrinos are about
the same: $W_{\nu} \sim W_p \sim W_{\rm GRB}$. Then the total number
of neutrinos with energy $E$ per burst is 
$$
N_{\nu}(E) \sim \frac{W_{\rm GRB}}{\ln E_{\rm max}/E_{\rm min}}E^{-2},
$$
where $E_{\rm min}$ and $E_{\rm max}$ are minimum and maximum acceleration
energy, respectively. Now one can write the flux of neutrinos from a
single GRB, expressing it through neutrino fluence $S_{\nu}$:
$$
F_{\nu_{\mu}+\bar{\nu}_{\mu}}(>E)=\frac{1}{3}~ 
\frac{S_{\nu}}{E\ln (E_{\rm max}/E_{\rm min})}.
$$
For $S_{\nu} \leq S_{\rm GRB}^{\rm max}= 1\times 10^{-4}~ {\rm erg/cm}^2$, 
the number of muons per GRB burst in ICECUBE is 
$N_{\mu} < 0.1$ for $E_{\mu} \geq 1$~TeV.

The diffuse flux is estimated in identical way through the local neutrino 
emissivity ${\cal L}_{\nu}(0)$ and evolutionary factor $k_{\rm evol}$ 
$$
J_{\nu_\mu+\bar{\nu}_{\mu}}(E)=\frac{1}{3}\frac{cH_0^{-1}}{4\pi}
\frac{{\cal L}_{\nu}(0)}{E^2\ln E_{\rm max}/E_{\rm min}} k_{\rm evol},
$$
where evolutionary factor is given by 
$$
k_{\rm evol}= \int_0^{z_{\rm max}}\frac{dz}{(1+z)^2}\frac{f_{\rm evol}(z)}
{\sqrt{(1+z)^3\Omega_m+\Lambda}}.
$$
For the evolutionary function $f_{\rm evol}(z)$ we take the case
of strong star formation evolution from \cite{wick}, which results in 
$k_{\rm evol}=7.0$ as a maximum value. For the emissivity we use   
$$
{\cal L}_{\nu}(0) \leq {\cal L}_{\rm GRB}(0)= 0.6\times 10^{43}~
{\rm erg}/{\rm Mpc}^3{\rm yr},
$$
where the local GRB emissivity is taken from \cite{schmidt}. 
As a result we obtain $E^2J_{\nu_\mu+\bar{\nu}_{\mu}}(E)=
3.9\times 10^{-10}$~GeV/cm$^2$ s sr, i.e. one order a magnitude lower 
than in \cite{waxman} (see Fig. \ref{limits}) (note that in \cite{waxman} 
the emissivity is by factor of 3 larger). Diffuse neutrino flux from
GRBs can be only marginally detected by ICECUBE , but time and
position correlations with GRBs make reliable even 1 - 3 detected
events.
\section{Non-accelerator neutrino sources} \label{non-acc}
\begin{figure}[t]
\epsfig{file=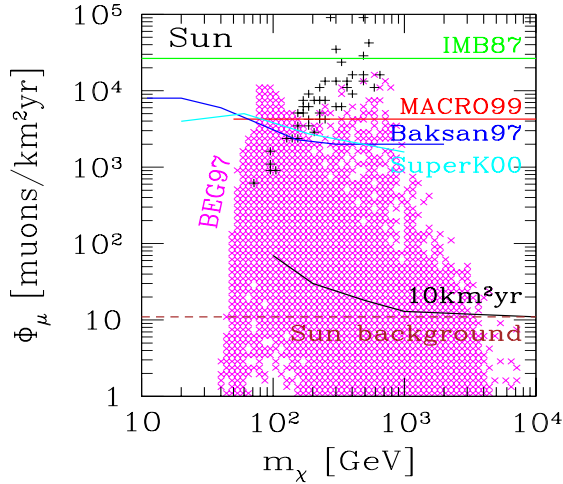,height=6.5cm,width=7.5cm}
\vspace{-14mm}
\caption{
Underground muon flux produced by neutralino annihilation in the 
Sun, from ref. \cite{gondolo}. The points correspond to different
supersymmetric parameters, 
$m_{\chi}$ is the neutralino mass. 
The upper curves give the upper limits from various
experiments. Curve ``Sun background'' corresponds to the neutrino flux 
induced by CR interacting with the Sun matter. The region above the curve 
10~km$^2$yr is accessible for a detector with the given exposure, e.g
for the ICECUBE detector. 
}
\vspace*{-5mm}
\label{neutralino}
\end{figure}

These sources include objects with annihilation of DM (the Sun, Earth, 
cores of the galaxies), objects with the decays of superheavy 
DM particles (galactic halos) and topological defects. In the last two cases  
neutrinos are produced in the decays of superheavy particles with the 
masses up to $M_{\rm GUT} \sim 10^{16}$~GeV. A particle decays to 
virtual particles, partons, which are cascading due to QCD
interaction, and at the confinement radius cascade partons are
converted to hadrons, most of which are pions. Neutrinos are produced
in pion decays with spectrum which can be approximately described at 
highest energies as $dE/E^2$. \\*[2mm]  
{\em Neutralino annihilation in the Sun and Earth}.\\*[1mm]
Neutralino is the best motivated DM particle. Crossing the 
Sun or Earth neutralino can loose its energy in collisions with 
nuclei and diminish its velocity below the escape velocity. If it
happens, the neutralino is gravitationally trapped in the object, and  
loosing further their energies, neutralinos are accumulated in the
center of a celestial body. Annihilating there they produce pions 
and then neutrinos. The process of annihilation strongly depends on 
neutralino mass and composition (mixture of basic fields: zino, bino
and two higgsinos). Fig.~\ref{neutralino} shows the neutrino-induced 
underground muon flux from annihilation of neutralinos in the Sun.
The Sun background is caused by neutrinos produced by cosmic rays 
bombarding the Sun.\\
{\em Topological defects} (TDs) are produced in early universe due to 
symmetry breaking accompanied by the  phase transitions.
In many cases TDs become unstable and decompose to constituent fields,
superheavy gauge and Higgs bosons (X-particles), which then decay
producing UHE neutrinos (see \cite{hill}, and 
\cite{berez99} for a review)\\*[1mm]
{\em Monopoles} (M) are produced due to $G \to H\times U(1)$ symmetry 
breaking. $M\bar{M}$ annihilation can produce QCD cascade and
neutrinos. Scenarios with the free monopole annihilation and with the bound 
$M\bar{M}$ systems (monopolonia) are disfavored for production 
of large neutrino fluxes.\\*[1mm]
{\em Ordinary strings} are produced by $U(1)$ symmetry breaking.  
There are several mechanisms by which ordinary strings can produce
HE neutrinos: collapse of the string loops, self-intersection,
annihilation of cusps, production and annihilation of tiny loops. 
In most cases produced neutrino fluxes are too small for detection.\\*[1mm]  
{\em Superconducting strings} can be powerful sources of neutrinos. 
In a wide class of elementary particle models strings behave like 
superconducting wires. Moving though cosmic magnetic fields, such
strings develop electric current. Superconducting strings produce
superheavy X-particles, when electric current in the string reaches
the critical value. X-particles escape from a string and decay.
This process is strongly enhanced near the {\em cusps}, the peculiar 
points on a string, where the velocity of an oscillating string reaches
the speed of light. The cusp emission of superheavy X-particles 
provides a realistic source of UHE neutrinos. \\*[1mm]
{\em Monopoles connected by strings} are produced in the 
$G \to H\times U(1) \to H \times Z_n$ sequence of symmetry breaking. 
At the first symmetry breaking monopoles are produced, at the second
one each monopole get attached to $n$ strings. Monopoles oscillate near
the equilibrium positions, generating UHE gauge bosons with  neutrinos 
produced at their decays.\\*[1mm]
{\em Necklaces} correspond to the case above when $n=2$, i.e. when 
each monopole is attached to two strings, and a loop reminds a necklace 
with monopoles playing the role of beads. In the process of evolution
the strings shrink due to gravitational radiation and $M \bar{M}$ 
pairs in the necklace inevitably annihilate. This model is most
plausible and well developed for UHE neutrino production. 
\vspace*{-5mm}
\section{Diffuse fluxes and upper limits}
In this section I will discuss the cascade and cosmic ray upper limits,
and compare them with the model calculations of diffuse fluxes from 
various astrophysical sources. 
\subsection{Cascade upper limit}
The cascade upper limit on HE and UHE neutrino fluxes \cite{BS,book}
is provided due to e-m cascades initiated by HE photons or electrons  
which always accompany production of neutrinos. Colliding with 
the target low-energy photons, a primary photon or electron produce 
e-m cascade due to reactions $\gamma+\gamma_{\rm tar} \to e^++e^-$,
$e+\gamma_{\rm tar} \to e'+\gamma'$, etc.  The standard case is given
by production of HE neutrinos in extragalactic space, and the cascade 
develops due to collisions with CMB photons ($\gamma_{\rm tar}= 
\gamma_{\rm CMB}$). In case neutrino production occurs in a galaxy,
the accompanying photon can either freely escapes from a galaxy
and produce cascade in extragalactic space, or produce cascade on 
the background radiation (e.g. infra-red) inside the galaxy. In the 
latter case the galaxy should be transparent for the cascade photons 
in the range 10~ MeV - 100~GeV. 

The spectrum of the cascade photons is calculated \cite{BS,book}:
in low energy part it is $\propto E^{-3/2}$, at high energies  
$\propto E^{-2}$ with a cutoff at some energy $\epsilon_{\gamma}$.
The energy of transition between two regimes is given approximately by
$\epsilon_c \approx (\epsilon_t /3)(\epsilon_{\gamma} /m_e)^2$, 
where $\epsilon_t$ is the mean energy of the target photon. 
In case the cascade develops in extragalactic space
$\epsilon_t=6.35\times 10^{-4}$~eV,
$\epsilon_{\gamma} \sim 100$~GeV (absorption on optical radiation),
and $\epsilon_c \sim 8$~MeV. The cascade spectrum is very close 
to the EGRET observations in the range 3~MeV - 100~GeV \cite{EGRET}.  
The observed energy density in this range is 
$\omega_{\rm EGRET} \approx (2 - 3)\times 10^{-6}$~eV/cm$^3$. 
The upper limit on HE neutrino flux $J_{\nu}(>E)$ (sum of all flavors) 
is given by chain of the following inequalities  
$$
\omega_{\rm cas}>\frac{4\pi}{c}\int_E^{\infty}EJ_{\nu}(E)dE>
\frac{4\pi}{c}E\int_E^{\infty}J_{\nu}(E)dE\equiv
$$
$$
\frac{4\pi}{c}EJ_{\nu}(>E),
$$
which in terms of the differential neutrino spectrum $J_{\nu}(E)$ gives
\begin{equation}
E^2 J_{\nu}(E) < \frac{c}{4\pi}\omega_{\rm cas },~~ {\rm with}~
\omega_{\rm cas} <\omega_{\rm EGRET}
\label{cas-rig}
\end{equation}

Eq. (\ref{cas-rig}) gives the {\em rigorous} upper limit on the neutrino flux. 
It is valid for neutrino production\\
\begin{figure}[t]
\epsfig{file=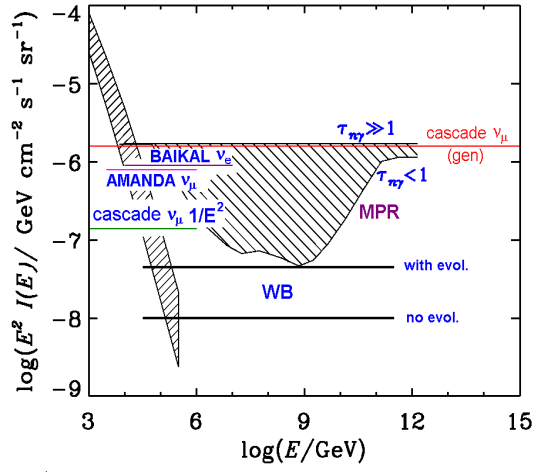,height=7cm,width=8cm}
\vspace{-15mm}
\caption{
Cascade and CR upper limits. The general cascade upper limit 
(\ref{cas-rig}) is shown by line ``cascade $\nu_{\mu}$'' 
with additional factor 1/3,
$E^{-2}$ cascade upper limit is given by line ``cascade 
$\nu_{\mu}~ 1/E^2$''. 
``BAIKAL $\nu_e$'' and ``AMANDA $\nu_{\mu}$'' 
are for Baikal and AMANDA upper limits, respectively. The 
Waxman-Bahcall upper limits are given by WB lines with and without 
evolution. Mannheim-Protheroe-Rachen limits are marked by MPR with 
the optical depth for neutrons: $\tau_{n\gamma}\gg 1$ 
and $\tau_{n\gamma}< 1$. Atmospheric neutrinos are shown by the
hatched diagonal strip. 
}
\vspace{-3mm}
\label{limits}
\end{figure}
by HE protons, by TDs, by
annihilation and decays of superheavy particles, i.e. in all cases
when neutrinos are produced through decay of pions and kaons. It is
valid for production of neutrinos in extragalactic space and in
galaxies, if they are transparent for the cascade photons. It holds
for arbitrary neutrino spectrum falling down with energy. If one assumes 
some specific shape of neutrino spectrum, the cascade limit becomes stronger.   
For $E^{-2}$ generation spectrum one immediately obtains 
\begin{equation}
E^2J_i(E) \leq \frac{1}{3} \frac{c}{4\pi}\frac{\omega_{\rm cas}}
{\ln (E_{\rm max}/E_{\rm min})},
\label{cas-E2}
\end{equation}
where $i=\nu_{\mu}+\bar{\nu}_{\mu}$ or $i=\nu_e+\bar{\nu}_e$.\\
AMANDA and Baikal collaborations in fact assume $E^{-2}$ neutrino 
spectrum to obtain 
observational upper limits. In Fig.~\ref{limits} we plot both 
the general cascade upper limit (\ref{cas-rig}) and $E^{-2}$ cascade
upper limit (\ref{cas-E2}), both for $\nu_{\mu}+\bar{\nu}_{\mu}$ 
neutrino flavors. For the $E^{-2}$ cascade upper limit, the AMANDA-B10
parameters $E_{\rm min}= 6$~TeV and $E_{\rm max}= 1000$~TeV
\cite{AMANDA} are used. Baikal upper limit is taken from \cite{Baikal}.
\\*[1mm]
{\em AMANDA and Baikal did not reach yet the cascade upper limit} 
(see Fig.~\ref{limits}).\\*[-7mm]
\subsection{Cosmic Ray (CR) upper limits}
Cosmic rays leaking out of HE neutrino sources should not exceed the
observed flux: this is the essence of CR upper limits on the 
neutrino flux.\\*[2mm]
{\em The Waxman-Bahcall upper limit\cite{WB1}}.\\*[1mm]
The neutrino sources which produce UHECR with energies $10^{19} - 10^{21}$~eV
are considered.  It is assumed that neutrinos are produced by protons in
$p\gamma$ collisions in a source, and that protons are loosing the fraction 
of energy $\epsilon <1$. The generation spectrum is assumed $\propto 1/E^2$.  
Escaping protons should not exceed the
observed flux of UHECR and it imposes the upper limit on neutrino flux. 
These upper limits are plotted in Fig.~\ref{limits} by the WB lines 
for evolutionary and non-evolutionary cases. 

Note, that this limit is not valid for sources with maximum energy 
of accelerated protons $E_{\rm max}\leq 1\times 10^{19}$~eV, for all sources
where protons are confined  (e.g. clusters of galaxies
and the AGN core models \cite{stecker}), and for all non-accelerator sources 
discussed above. \\*[2mm]
{\em The Mannheim-Protheroe-Rachen  upper limits.}\\*[1mm]
The CR upper limit can be relaxed taking into account the absorption 
of HE protons in a source. In the work  \cite{MPR} the exit of protons 
from a source is provided by production and escape of the neutrons with 
their consequent decays to the protons.  This upper limit is
constructed mainly for AGN, assuming various optical depth 
$\tau_{n\gamma}$ for escaping protons. In contrast to WB limit, 
the maximum energy of accelerated protons is allowed in a wide range 
$10^6 - 3\times 10^{13}$~GeV. The upper bound on extragalactic proton 
flux for this energy interval is extracted from observations. The
spectrum of accelerated protons is assumed $\propto 1/E^2$, but 
the resulting proton spectrum at observation is evaluated. 
\begin{figure}[t]
\epsfig{file=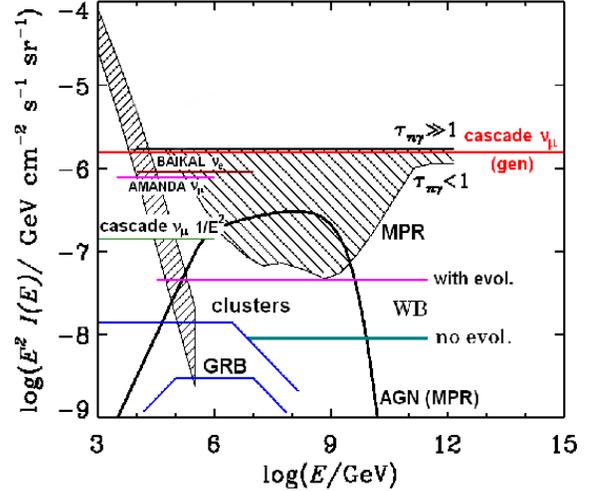,height=6.5cm,width=7.5cm}
\vspace{-15mm}
\caption{
Diffuse fluxes from AGN jets, GRBs and galaxy clusters, compared with 
upper bounds.  
}
\vspace*{-7mm}
\label{agn}
\end{figure}
In Fig.~\ref{limits} the MPR upper limits are shown for two extreme 
cases $\tau_{n\gamma}<1$ (neutron transparent sources) and 
$\tau_{n\gamma} \gg 1$ (neutron non-transparent sources).\\*[1mm] 
CR upper limits are valid for AGN and GRBs. They are not valid for 
TDs, annihilation and decay of DM particles and hidden sources \cite{WB2}.  
They are also not valid for such conservative accelerator sources as 
galaxy clusters. Such long list of exclusions contradicts the standard 
definition of upper limit in physics. It deserves more the name
landmark for future HE neutrino detectors, especially in the case of the WB
limit.
\\*[2mm]
In Fig.~\ref{agn} diffuse fluxes  from AGN jets, GRBs and galaxy
clusters are shown in comparison with cascade upper limits and CR
benchmarks. Note, that diffuse flux from galaxy clusters 
(the case without evolution) exceeds the corresponding WB limit. 
\section{SuperGZK neutrinos} 
Soon after prediction of the GZK cutoff, it was noticed 
that this phenomenon is accompanied by a flux of extremely high 
energy neutrinos, 
\begin{figure}[t]
\epsfig{file=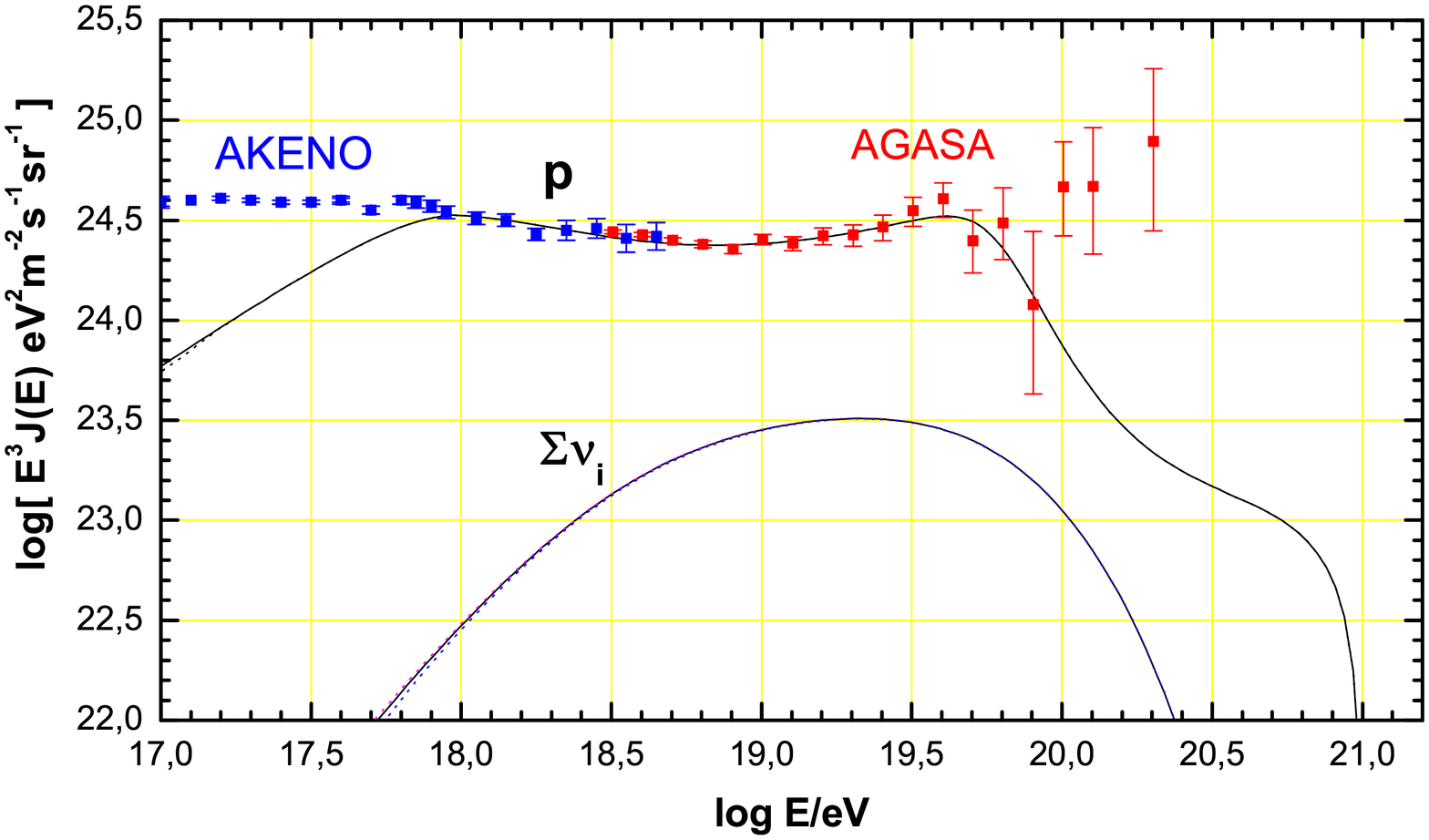,height=5.5cm,width=7.5cm}
\vspace{-15mm}
\caption{
Diffuse proton and cosmogenic neutrino fluxes in the AGN model \cite{BGG},
without cosmological evolution and with 
$E_{\rm max}=1\times 10^{21}$~eV. The generation spectrum index 
is taken $\gamma_g=2.7$ to provide the best fit 
of the Akeno-AGASA data by the proton spectrum (curve p).  
}
\vspace*{-8mm}
\label{cosmog-low}
\end{figure}
which in some models exceeds the flux of parent protons \cite{BZ}. 
For detection of these neutrinos the horizontal EAS have been proposed 
\cite{BS}.  The boost to this field has been  recently given by 
the projects
of observation of superGZK neutrinos from the space \cite{euso} and 
by the first results of observation of   
radio signal from the neutrino-induced cascades in the 
ice \cite{ice}  and in the Moon \cite{moon}.  In all these detectors 
the threshold is very high, and these methods are mostly 
effective for superGZK neutrinos with energies 
$E \gsim 1\times 10^{20}$~eV.    

In what sources these tremendous energies are possible? 

The accelerator sources are disfavored, if one limits himself by the 
shock acceleration. Neutrino energy $E_{\nu}> 1\times 10^{20}$~eV 
implies proton energy $E_p \sim 20 E_{\nu} > 2\times 10^{21}$~eV. 
Such energy is only marginally possible for non-relativistic
shocks in AGN 
jets,  and it is too high for ultrarelativistic shocks \cite{GA}. However, 
there are many proposals of plasma mechanisms of acceleration 
with $E_{\rm max}$ much higher than $1\times 10^{21}$~eV, see 
e.g. \cite{plasma}.  One may optimistically expect that 
$E_{\rm max} \sim 1\times 10^{22}$~eV is not the excluded possibility. 

TDs and SHDM particles naturally provides very high energies up to 
$E_{\rm GUT} \sim 10^{16}$~GeV. 
\subsection{Cosmogenic neutrinos}
These neutrinos are produced by UHECR interacting with CMB: 
$p+\gamma_{\rm CMB} \to \pi^{\pm} \to {\rm neutrinos}$. 
The cosmogenic neutrino flux can be approximately evaluated using
the unmodified CR flux $J_p^{\rm unm}(E)$, i.e. one where 
only adiabatic energy losses (red shift) are taken into account. 
\begin{figure}[t]
\epsfig{file=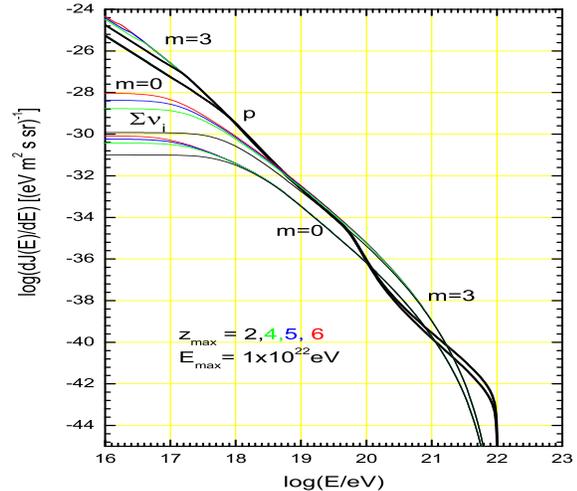,height=6.5cm,width=7.5cm}
\vspace{-15mm}
\caption{
Diffuse cosmogenic neutrino fluxes in the AGN model \cite{BGG} with
and without evolution. The evolution is described by factor  $(1+z)^m$ up 
$z_{\rm max}$. The maximal energy is $E_{\rm max}=1\times 10^{22}$~eV
in all cases. The non-evolutionary case is shown by the curves $m=0$.
The proton spectrum, given by curve p, fits well all observations. 
}
\vspace*{-6mm}
\label{cosmog-high}
\end{figure}
Neutrino flux at $z=0$ is given by \cite{BZ}
\begin{equation}
J_{\nu}(E)=\frac{2}{3}3\left (\frac{E_{\nu}}{E_p}\right )^{\gamma_g -1}
\frac{1}{1-\alpha^{\gamma_g-1}}J_p^{\rm unm}(E),
\label{estimate}
\end{equation}
where 2/3 is a fraction of charged pions produced by very high energy
proton, $E_{\nu}/E_p \approx 0.05$ is a fraction of proton energy
transferred to neutrino, and $\alpha$ is a fraction of energy lost 
by proton in  $p\gamma$ collision ($\alpha$ varies from 0.22 in 
$\Delta$-resonance to 0.5 at extremely high energies). The term with 
$\alpha$ in Eq.(\ref{estimate}) takes into account the multiple
collisions of proton with CMB photons, especially important in
the evolutionary case. At low energies neutrino flux becomes flat.      
In evolutionary models 
the unmodified proton flux and neutrino flux can be much higher than 
UHECR flux calculated with all energy losses included. 
The recent calculations of cosmogenic neutrinos \cite{cosmog-nu} 
use the normalization to the observed UHECR flux. The neutrino fluxes
are small in non-evolutionary models with 
$E_{\rm max} \leq 1\times 10^{21}$~eV and large 
in the models with strong cosmological evolutions and with 
$E_{\rm max} \geq 1\times 10^{22}$~eV.   
In Fig.~\ref{cosmog-low} we present the neutrino flux produced in 
the non-evolutionary AGN model \cite{BGG} with 
$E_{\rm max}=1\times 10^{21}$~eV and with generation spectrum index 
$\gamma_g=2.7$. This model describes well the UHECR spectra observed by all 
existing detectors, except the highest energy AGASA events 
(see Fig.~\ref{cosmog-low}). The calculated neutrino spectrum \cite{BGG1} 
gives 
the low limit on the cosmogenic neutrino flux imposed by the observed flux 
of UHECR and assumption $E_{\rm max} \leq 1\times 10^{21}$~eV. This
flux is smaller than the low limit found in the paper by Fodor et al 
\cite{cosmog-nu} and is essentially undetectable.
\begin{figure}[t]
\epsfig{file=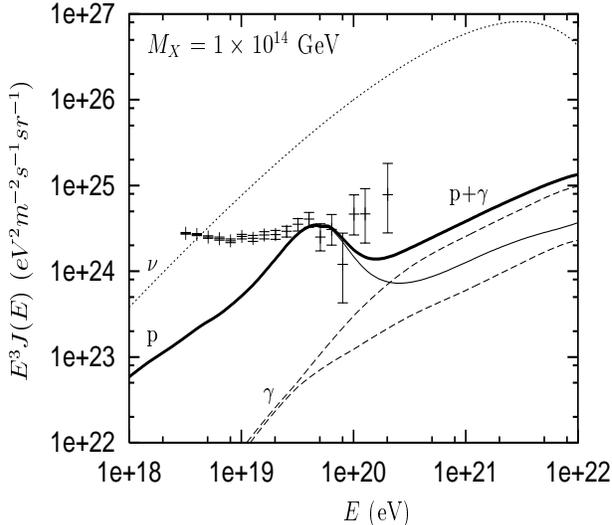,height=7cm,width=8cm}
\vspace{-15mm}
\caption{
Diffuse neutrino spectrum from necklaces for $M_X=1\times 10^{14}$~GeV
\cite{ABK}. The thick curve gives $p+\gamma$ flux normalized to the
AGASA data.   
}
\vspace*{-7mm}
\label{neckl}
\end{figure}
The large neutrino fluxes produced in the AGN models with  
$E_{\rm max} \leq 1\times 10^{22}$~eV are shown in
Fig.~\ref{cosmog-high} \cite{BGG1}. The calculated UHECR flux 
in all cases fits well the
observational data. Both the models with cosmological
evolution $(1+z)^m$ up to $z_{\rm max}$, and non-evolutionary models 
with m=0  were used. One can notice that evolutionary models predicts
much larger neutrino fluxes. They are detectable by future arrays. 

In case the UHECR primaries are extragalactic nuclei, the flux of superGZK
neutrinos is suppressed. The nuclei responsible for superGZK neutrinos  
are promptly photo-disintegrated to A nucleons, which produce neutrinos 
in photopion collisions. Thus, the energy of neutrino is 
$E_{\nu} \sim (0.05/A)E_N$, where $E_N$ is an energy of nucleus. 
Putting $E_{\nu}/E_N$ ratio in Eq.~(\ref{estimate}), and taking 
into account 3A neutrinos produced by one nucleus, we obtain the
additional suppression of ratio $J_{\nu}(E)/J_N^{\rm unm}(E)$ by 
factor $A^{2-\gamma_g}$. Besides, the energy of accelerated nucleus 
should be $E_N > 20A \times 10^{20}$~eV, e.g. $1.1\times 10^{22}$~eV
in case of iron. 

The extragalactic nuclei as UHECR primaries have another interesting
aspect connected with neutrinos of lower energies $E_{\nu} \lsim 10^{15}$~eV.  
In this energy region the neutrino production is dominated by neutron decays
with neutrons produced in photodissociation of UHE nuclei
\cite{nuclei}.
\subsection{Topological defects}
TDs as neutrino sources are described in Section \ref{non-acc}.
Neutrino fluxes from TDs are limited by the cascade upper limit 
(\ref{cas-rig}). Usually it excludes too large masses of X-particles, 
in decays of which neutrinos are produced. 
The diffuse neutrino flux from {\em necklaces} has been calculated in 
\cite{ABK}. The neutrino spectrum is shown in Fig.~\ref{neckl} for 
$M_X=1\times 10^{14}$~GeV, allowed by the cascade upper limit 
(see Fig.~\ref{mirr-nu}).
The spectrum is normalized to fit the AGASA data with help of protons 
and photons ($p+\gamma$ curve). One can see that in superGZK region 
the neutrino flux is two orders of magnitude higher than $p+\gamma$
flux. This model provide much higher neutrino energies than that 
of cosmogenic neutrinos and much higher fluxes.    
\subsection{Mirror neutrinos}
Mirror matter can be most powerful source of superGZK neutrinos not limited
by the usual cascade limit \cite{mirror}. 

Existence of mirror matter is based on the deep theoretical concept, which
was  introduced by Lee and Yang \cite{LY}, Landau \cite{Landau} and 
most notably
by Kobzarev, Okun and Pomeranchuk \cite{KOP}. Particle space is a
representation of the Poincare group. Since the space reflection 
$\vec{x} \to -\vec{x}$ and
time shift $t \to t+\Delta t$ 
commute as the coordinate transformations, the
corresponding inversion operator $I_s$ and the Hamiltonian $H$ must
commute, too: $[I_s,H]=0$. Because the parity operator $P$ does not commute 
with $H$ (i.e. parity is not conserved) Lee and Yang suggested that 
$I_s=P\cdot R$, where the operator $R$ generates the mirror particle 
space, and thus $I_s$ transfers 
the left states of ordinary particles into right states of the mirror 
particles and vise versa. In fact, the assumption of Landau is
similar: one may say that he assumed $R=C$. 

The mirror particles have interactions identical 
to the ordinary particles, but these two sectors interact with each other
only gravitationally \cite{KOP}. Gravitational interaction mixes the visible 
and mirror neutrino states, and thus causes the oscillation between
them.  
\begin{figure}[t]
\epsfig{file=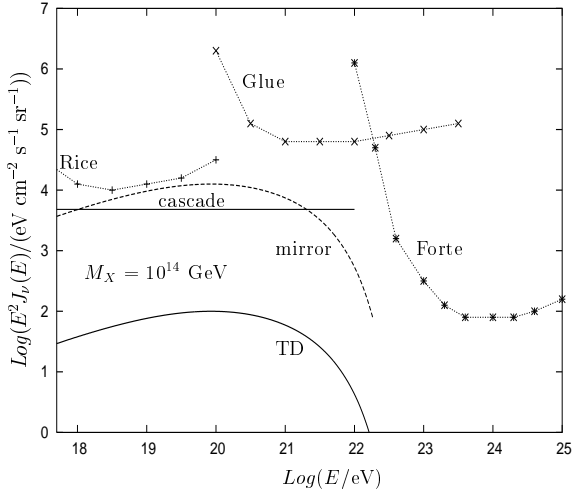,height=6.5cm,width=7.5cm}
\vspace{-15mm}
\caption{
Diffuse flux of the visible neutrinos from mirror necklaces 
with $M_X=1\times 10^{14}$~GeV. The flux is limited by observations 
of RICE, GLUE and FORTE. Note, that neutrino flux exceeds the
general cascade upper limit (\ref{cas-rig}). TD curve gives the flux
from the ordinary necklaces.
}
\vspace*{-7mm}
\label{mirr-nu}
\end{figure}
A cosmological scenario must provide the suppression of the mirror
matter and in particular the density of mirror neutrinos at the epoch 
of nucleosynthesis. It can be obtained in the two-inflaton model 
\cite{mirror}. The rolling of two inlatons to minimum of the potential
is not synchronised, and when the mirror inflaton reaches minimum, the ordinary
inflaton continues its rolling, inflating thus the mirror matter produced
by the mirror inflaton. While mirror matter density is suppressed, the mirror
topological defects can strongly dominate \cite{mirror}. Mirror TDs 
copiously produce mirror neutrinos with extremely high
energies typical for TDs, and they are not accompanied by
any visible particles. Therefore, the upper limits on HE mirror 
neutrinos in our world do not exist. All HE mirror particles 
produced by mirror TDs are sterile for us, interacting with
ordinary matter only gravitationally, and only mirror neutrinos 
can be efficiently converted into ordinary ones due to oscillations. 
The only (weak) upper limit comes from the resonant
interaction  of converted neutrinos with DM neutrinos: 
$\nu+\bar{\nu}_{\rm DM}\to Z^0$ \cite{mirror}.
The strongest limit on the fluxes of superGZK neutrinos are given by 
radio observations \cite{ice} and \cite{moon}. In Fig.~\ref{mirr-nu} 
we present the flux of visible neutrinos from the mirror necklaces 
with $M_X=1\times 10^{14}$~GeV, limited by radio observations.
\section{Conclusions}
High energy neutrino astronomy with the future ice/underwater 
1~km$^3$ detectors will reach the decisive stage in searching for
neutrino sources and diffuse fluxes.\\*[1mm]
At present with AMANDA and Baikal the most rigorous {\em cascade 
upper limit} on the diffuse flux is almost reached. To demonstrate 
existence of this limit is an important goal, which will prove 
the absence of exotic physics such as e.g. large diffuse fluxes from the
hidden sources. \\*[1mm]
Detection of {\em accelerator sources} (and/or diffuse fluxes from
them) such as AGN and GRBs will clarify the nature of these sources 
and acceleration mechanisms.\\*[1mm]
Detection of {\em non-accelerator sources} might result in discovery
of ``new'' physics: (i) Supersymmetry and Dark matter in case of HE
neutrinos from the Sun and Earth, (ii) Topological Defects (diffuse
flux is higher than {\em CR upper limits} or presence of neutrinos   
with energies $E_{\nu} > 10^{21} - 10^{22}$~eV), and (iii) Mirror
Matter (diffuse flux is higher than {\em cascade upper limit}.\\*[1mm]
Detection of superGZK neutrinos will be the real break-through either 
in acceleration physics or in physics beyond the Standard Model 
(Topological Defects, Superheavy Dark Matter, Mirror Matter). 


\end{document}